\documentstyle[prl,aps,epsf]{revtex}
\draft
\begin{document}
\twocolumn[\hsize\textwidth\columnwidth\hsize\csname @twocolumnfalse\endcsname

\title{Dynamic Ordering and 
Transverse Depinning of a Driven Elastic String in a Disordered Media}
\author{C. Reichhardt$^{\rm (1)}$ and C. J. Olson$^{\rm (2)}$}
\address{$^{\rm (1)}$CNLS and Applied Theoretical and Computational Physics Division,} 
\address{$^{\rm (2)}$ Theoretical and Applied Physics Divisions,}
\address{Los Alamos National Laboratory, Los Alamos, NM 87545}

\date{\today}
\maketitle
\begin{abstract}
We examine the dynamics of an elastic string interacting with
quenched disorder driven perpendicular and parallel to the string.
We show that the string is the most disordered at the depinning transition
but with increasing drive partial ordering is regained.
For low drives the noise power is high and we observe a $1/f^2$ noise
signature crossing over to a white noise character with low power at 
higher drives. 
For the parallel driven moving string there is a finite transverse critical 
depinning force with the depinning  transition occuring by the formation of
running kinks. 
\end{abstract}
\vspace{-0.1in}
\vspace{-0.3in}

\vskip2pc]
\narrowtext

The dynamics of driven elastic media interacting with 
quenched disorder occurs 
in a wide variety of physical systems which include 
magnetic domain wall motion \cite{Bertotti1},
nonequilibrium growth \cite{Barabasi2},
models of friction \cite{friction3}, vortex lattice motion in 
superconductors \cite{Blatter4}, 
and charge density waves \cite{Balent5}.
Recently 
an intense interest in driven elastic media with quenched disorder 
has been directed at {\it dynamical ordering} where at low applied drives
the system is in a highly disordered pinned state and at a critical
drive a depining into a highly disordered moving state occurs, while at 
high drives the effect of the disorder is reduced and
the system can regain a considerable amount of order. 
Particular systems where this reordering 
behavior has been studied extensively 
experimentally 
\cite{Higgins6,Yaron7,Pardo8,Kes9}, 
theoretically \cite{Koshelev10,Giamarchi11,Balents12,Sheidl13}, 
and numerically \cite{Koshelev10,Moon14,Berlinsky15,Olson16,Transverse17} 
include vortices
in superconductors
as well as
charge-density wave (CDW) systems. 

In the vortex system if there is no quenched disorder 
the vortices form an ordered crystalline state; 
however, in most samples there are defects which 
attract and pin vortices, disordering the vortex
lattice.  
When a driving force is applied,
the initial depinning transition can be 
plastic, with 
certain regions of mobile vortices tearing past pinned regions
and a large number of defects generated so that the system has only
a liquid like structure.  
For higher drives the vortices regain order; however, 
the pinning can still affect the structure so the vortices 
can have a partial anisotropic ordering or smectic structure
\cite{Balents12}.
An open question is
whether the reordering transition is a true transition or a 
crossover.    

An interesting feature of the reordered moving state is that
a {\it transverse depinning threshold} can exist as originally proposed
in Ref. \cite{Giamarchi11}. 
Here, although the lattice is moving in the
longitudinal direction, the effect of the pinning is still present in the
transverse direction so that a finite transverse drive must be applied before
the lattice will move in the transverse direction.
Simulations \cite{Moon14,Transverse17} and 
experiments \cite{Kes9} 
on vortex lattices as well as simulations 
for friction models \cite{friction3} 
have found evidence for a finite transverse depinning threshold for
the longitudinal moving systems. 
Other theoretical work contends that at finite $T$ no 
true critical transverse depinning threshold exists;
instead, due to the high pinning barriers,
a pronounced {\it crossover} in the transverse IV curves may appear
\cite{Balents12}.
In the CDW context the effect of the quasi-particle current flowing
{\it perpendicular} to the CDW 2$k_F$ ordering wavevector was
also studied theoretically \cite{Leo18} and in experiment \cite{Markovic19}.
An intriguing question is whether the reordering transitions and
the transverse barrier   
seen in the vortex system occur in other types of
driven elastic media, such as
driven interfaces and polymers interacting with quenched disorder. 
An ideal way to model these systems 
in 2D is with a driven elastic
string interacting with quenched disorder.  

Previous numerical studies of driven elastic strings with disorder 
\cite{String120,Kaper21,Tang22,Makse23}
employed drives perpendicular to the string and
focused 
on the dynamics near the depinning threshold.  In this case
critical behavior is expected, and was observed in the form of broad 
distributions of avalanche sizes as 
well as scaling of the velocity near depinning, $v = (f - f_{c})^{\alpha}$,
as proposed by Fisher \cite{Fisher24}. 
The roughness of the string 
as measured by the roughening coefficient
exponent was $0.9$ to $1.1$ at depinning 
\cite{String120,Tang22} 
and $0.5$ at high drives \cite{String120}, 
indicating that
some ordering of the string was occurring as the drive increased. 

In this work we investigate the
continual evolution of the order
and noise signatures in driven strings for both parallel and
perpendicular drives.  
We also investigate the transverse depinning for the moving string. In 
the case of the perpendicular moving string we do not observe a transverse
barrier; however, for the parallel driven string a transverse depinning
threshold is observed which occurs through the formation of running 
kinks. For increasing longitudinal drives the transverse barrier decreases. 
Physical systems relevant to the perpendicular driven string include
domain walls and moving interfaces, while
systems 
similar to the parallel driven 
string include boundaries between sliding surfaces, 
polymers that are aligned with an applied drive in a random media,
or a single stream of fluid flowing down a rough surface. 


\begin{figure}
\center{
\epsfxsize=3.4in
\epsfbox{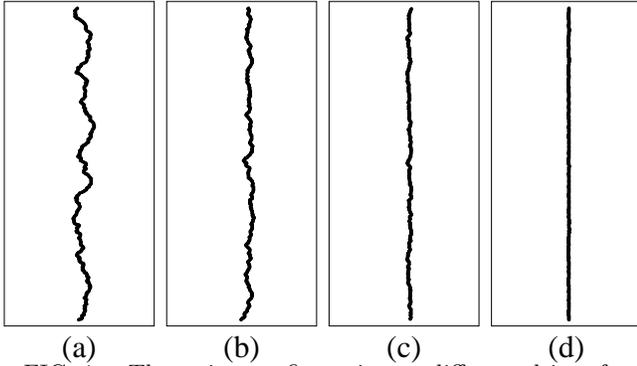}}
\caption{
The string configuration at different drives
for a system with driving in the $x$-direction 
perpendicular to the string with
a depinning threshold of $F_{dp} = 0.09$. 
(a) Right above depinning at $F_{d}/F_{dp} = 1.1$; 
(b) $F_{d}/F_{dp} = 1.38$; (c) $F_{d}/F_{dp} = 1.94$; (d) 
$F_{d}/F_{dp}=4.4$.} 
\label{fig1}
\end{figure}


We model an overdamped elastic string in (1 + 1) 
dimensions. The string is
composed of discrete particles 
connected by springs.  The system
has periodic boundary conditions in the $x$ and $y$-directions and the 
string
is connected periodically in the $y$-direction.
The equation of motion for each particle on the string is:
\begin{equation}
\gamma \frac{{\bf dr}}{dt} = \kappa {\bf f}_{s} + {\bf f}_{p} + {\bf F}_{d} \ .
\end{equation} 
Here ${\bf r}$ is the particle location, $\gamma=1$ is the damping term, 
$\kappa=5$ is the string elastic constant, 
${\bf f}_{s}={\bf r}_{i-1}-2{\bf r}_{i}+{\bf r}_{i+1}$.
is the spring force from the two neighboring
particles, ${\bf f}_{p}$ is the pinning force and ${\bf F}_{d}$ 
is the uniform applied
driving force.  The pinning is modeled as attractive parabolic traps 
scattered randomly through the sample.
The pinning force is
${\bf f}_{p} = -\sum_{i}^{N_{p}}(f_{p}/r_{p})
({\bf r}_{i}-{\bf r}_{k}^{(p)})
\Theta(r_{p} -  
|{\bf r}_{i} - {\bf r}_{k}^{(p)}|),$
where $\Theta$ 
is the Heaviside step function, ${\bf r}_{k}^{(p)}$ is the location
of pinning site $k$, $f_{p}=0.3$ is the maximum pinning force, and
$r_{p}=0.25$ is the pinning radius.
The results shown here are for strings containing $N=500$ 
to $2000$ particles,
interacting with $N_p=4.6 \times 10^5$ 
pinning sites in samples of size $160 \times 160$.
We consider two types of initial conditions: one where the string
is put down in its unstretched equilibrium position, and a second where
the string is annealed
at a finite temperature by the addition of a thermal kick. 
We find that when the applied drive is increased slowly enough, 
both methods give similar results. We apply a uniform force $F_{d}$  
in the $x$ direction for perpendicular driving or the $y$-direction for 
parallel driving. We increase the driving 
force in increments of 
$0.001$ and spend $30000$ simulation 
steps at each increment, measuring the 
average string velocity
$V_{x} = \sum_{i}^{N_{i}}{\hat {\bf x}}\cdot {\bf v}_{i}$,
and 
$V_{y} = \sum_{i}^{N_{i}}{\hat {\bf y}}\cdot {\bf v}_{i}$.
To quantify the order in the string we measure the
difference $\Delta L$ in the length of the string compared to the equilibrium
length, 
$\Delta L = \sum_{i}^{N_{i}}(|{\bf r}_{i} - {\bf r}_{i+1}| - r_{l})$,
where $r_{l}$ is the equilibrium length of each string segment. 
We also measure the
noise in the string velocity:
$S(\omega) = |\int V_{x}(t)e^{-i2\pi\omega t} dt|^2,$
and 

\begin{figure}
\center{
\epsfxsize=3.4in
\epsfbox{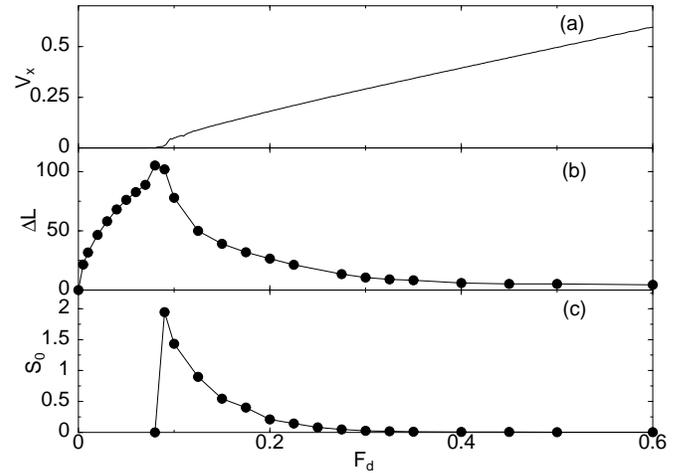}}
\caption{
(a) The string velocity $V_{x}$ versus $F_{d}$ for the system in
Fig.~1. The string is pinned for $F_{d} < 0.09$. (b) $\Delta L$ versus
$F_{d}$ showing that the string becomes disordered as $F_{dp}$
is approached from below, and 
gradually reorders for increasing $F_{d} > F_{dp}$.
(c) Noise power $S_{0}$ versus $F_{d}$.}
\label{fig2}
\end{figure}

\hspace{-13pt}
the noise
power 
in one frequency octave 
$S_{0} = \int_{\omega_{2}}^{\omega_{1}}S(\omega)$.

In Fig.~1(a-d) we show a 
series of snapshots of a perpendicular driven string.   
In Fig.~1(a) at a driving force 
$F_d/F_{dp}=1.1$ very close to the 
depinning transition $F_{dp}=0.09$,
the string is highly disordered. 
Above depinning, the motion of the string is jerky, occurring in pulses or 
avalanches, with portions of the string immobile while other portions
are moving as can be seen from the instantaneous velocity of individual beads.
In Fig.~1(b) 
and (c) for $F_{d}/F_{dp} = 1.25$ and $1.75$, 
the string becomes less winding. 
Finally in Fig.~1(d), for $F_{d}/F_{dp} = 4.0$ the string 
is almost completely ordered again. 

To quantify the evolution of the driven string from Fig.~1, we plot in 
Fig.~2(a) $F_{d}$ versus $<V_{x}>$ showing that the string is pinned
for $F_{d} < 0.09$. 
In this case the string was initially placed in its equilibrium position. 
In Fig.~2(b) we plot $\Delta L$ which shows that 
the string is aligned 
for $F_{d} = 0$ but gradually disorders as $F_{dp}$
is approached from below.  $\Delta L$ reaches a maximum
at depinning. For increasing 
$F_{d} > F_{dp}$, $\Delta L$ gradually decreases
as the string orders. In Fig.~2(c) 
the reordering can also be seen in the
noise power $S_{0}$ 
which is maximum at 
depinning and again 
decreases for increasing 
$F_{d}$. The decay of $S_{0}$ fits well to an exponential form
where for $F_{d} > F_{dp}$, 
$S_{0}(F_{d}) \propto \exp( -(F_{d} - F_{dp}))$.  
We have also conducted simulations where we
decrease $F_{d}$ after a 
ramp up, and we do not observe any 
hysteresis in the ordering $\Delta L$ or the noise power $S_0$. 

In Fig.~3 we show noise spectra for the same system.
Right above depinning, at $F_{d}/F_{dp}=1.1$,
$S(\omega)$ shows a $1/f^{2}$ characteristic,
as seen in Fig. 3(a).  For increasing drives, 
the string begins to move continuously and reorder.  As a result, a 
characteristic
time scale forms, 
the $1/f^{2}$ shape is lost and the 
higher frequencies are reduced 
in power, as shown in Fig. 3(b) 
for
$F_{d}/F_{dp}=4.4$. Noise 

\begin{figure}
\center{
\epsfxsize=3.4in
\epsfbox{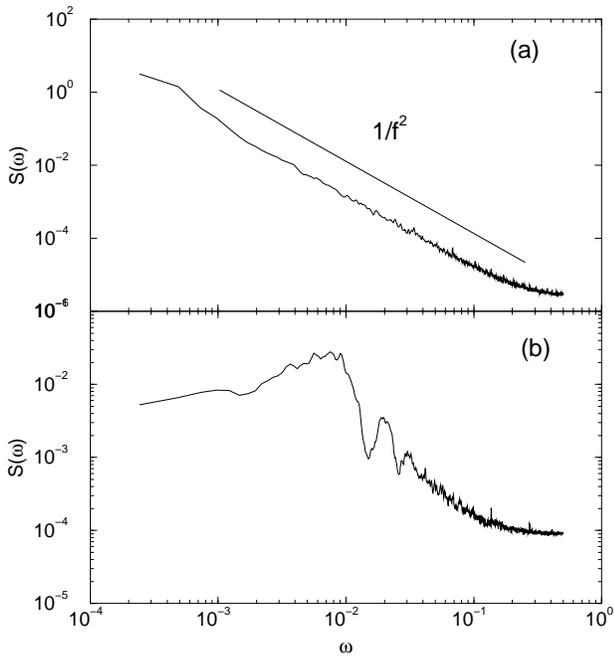}}
\caption{
(a) Noise spectra $S( \omega )$ for the system in Fig.~1 for 
$F_{d}/F_{dp} = 1.1$ showing a $1/f^{2}$ signal. (b) 
$S( \omega )$ for $F_{d}/F_{dp} = 4.4$ showing 
a characteristic time scale.} 
\label{fig3}
\end{figure}

\hspace{-13pt}
spectra for vortices have shown similar features
in both experiments \cite{Shobo25} and simulations \cite{Olson16}. 
In the vortex simulations the $1/f^2$ high noise region corresponds to the 
disordered plastic flow state and the low noise power region
corresponds to the reordered elastic flow state.

We have repeated the same type of simulations shown in Figs.~1-3 for the
parallel driven string and observe the same general features. 
For the parallel driven string, the depinning threshold is 
lower and we observe a clear washboard signal for the high drive case. 
We have carried out a  series of simulations for varied $f_{p}$,
$\kappa$, and $N_{p}$.  In each case, the depinning threshold changes;
however, the same general features of the reordering are observed. 

To test the predictions in Ref. \cite{Giamarchi11} of a critical transverse 
threshold for the driven string, we have conducted 
a
series of simulations where we fix the
driving force at a constant value after the string is in motion.  We
then slowly apply an 
additional driving force in the direction transverse to the initial 
longitudinal motion. For the perpendicular driven string the applied 
transverse drive is 
along the string and for the parallel driven string the 
applied transverse
drive is perpendicular to the string. For the perpendicular driven 
string we
do not observe a 
transverse barrier for the moving string. 
For a parallel driven string, in 
Fig.~4(a) we show the transverse driving force versus the transverse string
velocity indicating that 
for the case of the parallel moving string 
there is a
finite transverse pinning threshold.  
We note that for the 
moving vortex systems a finite transverse barrier occurs only 
when the vortices are moving in well defined channels 
and the transverse
drive is perpendicular to these 

\begin{figure}
\center{
\epsfxsize=3.4in
\epsfbox{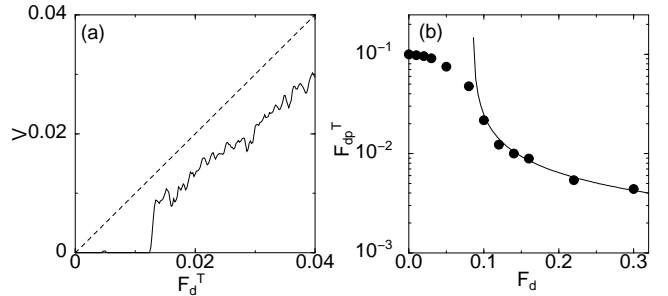}}
\caption{ 
(a) The transverse string velocity $V$ versus the transverse drive 
$F_{d}^{T}$ for a parallel driven string (lower curve). 
The parallel string drive is 
$F_{d}/F_{dp} = 1.22$. A finite transverse depinning threshold is visible. A
plot of $V$ for a system with no quenched disorder 
is also shown (dashed curve). (b) The transverse depinning threshold
$F_{dp}^{T}$ versus longitudinal drive $F_{d}$.  The solid line indicates
a fit to $F_{dp}^{T} \propto (F_{d}-F_{dp})^{-2/3}$}
\label{fig4}
\end{figure}

\hspace{-13pt}
channels. The parallel moving string 
may be viewed as a single moving channel. 
In Fig.~4(b) we show that
the transverse barrier decreases with increasing longitudinal velocity. 
In the vortex system as
the longitudinal drive increases the channels straighten out and the
transverse barrier decreases. In the parallel moving string the
decrease in the transverse pinning threshold also corresponds to the
string 
reordering for the higher longitudinal drives. 
Above the parallel depinning transition, $F_{d}>F_{dp}$, 
the data follows a power law form, 
$F_{dp}^{T} \propto (F_{d}-F_{dp})^{-2/3}$.  In a CDW system an exponential
decay would be expected \cite{Leo18}, but for the 
driven string system a power law decay is
predicted instead \cite{privateLeo26}.

We have also 
investigated the dynamics of the transverse depinning 
for the parallel driven string, as indicated in 
Fig.~5, which shows a series of snapshots of the string just 
above the transverse depinning threshold. 
The transverse depinning transition is very distinct from
the longitudinal depinning of the
perpendicular driven string shown in Fig.~1.  The 
transversely driven parallel moving string 
does not become more disordered as observed at the depinning transition of the
strictly perpendicular driven string.  Instead, the 
transverse depinning 
occurs through the formation of a running kink that moves in the direction
of the longitudinal drive as seen in Fig.~5. Although one might expect 
a kink and an anti-kink to 
form, we observed that below the
kink the string is positioned at a slight angle. The kink moves
along the string through the periodic boundary conditions. As the
transverse drive is increased more kinks appear and the kinks begin 
to move more rapidly.
Similar kink motion has been observed in interfaces in sliding
friction systems 
\cite{private27}.
The moving kinks themselves can show intricate dynamics.  
We have observed that these kinks can become pinned, and that
another kink can form elsewhere and 
later collide with the pinned kink 
to form a larger kink.   

In conclusion we have numerically investigated the 
dynamics of 
perpendicular and parallel driven strings in 

\begin{figure}
\center{
\epsfxsize=3.4in
\epsfbox{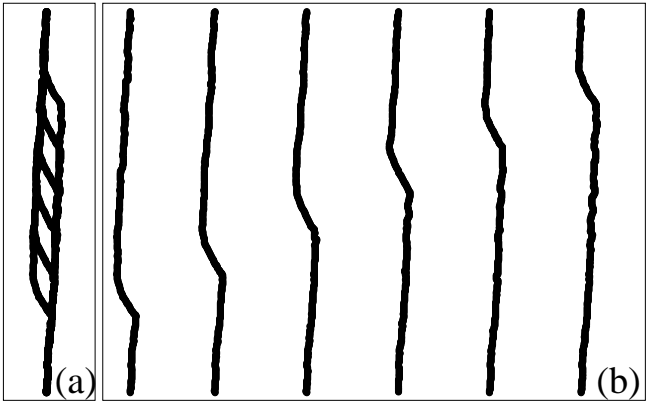}}
\caption{ 
(a) A series of snapshots of a parallel driven string 
as a function of time 
with the longitudinal drive in the $y$-direction and the transverse 
drive in the $x$-direction, 
for a system
with $F_{d} = 0.12$, just above the transverse depinning threshold.
The transverse depinning occurs via formation of a running kink that
moves 
in the direction of the longitudinal drive.
(b) The same snapshots offset in the $x$ direction.}
\label{fig5}
\end{figure} 

\hspace{-13pt}
quenched disorder. We find 
that the string is most disordered at the depinning transition but
recovers order 
for increasing drive as measured by the difference between the 
length of the string and the equilibrium length.
This reordering is observable in the noise 
characteristic.  Broad $1/f^2$ noise is
observed at depinning, with a gradual reduction in the noise power
for increasing drive and a shift to a whiter noise characteristic. For the
parallel driven moving string we find that there is a finite transverse
critical depinning threshold $F_{dp}^{T}$, indicating that 
the theory of Ref. \cite{Giamarchi11} regarding the transverse barrier
in a very different system still appears to apply to this system. 
We see a decrease of $F_{dp}^{T}$
with increasing longitudinal drive, as predicted in Ref. \cite{Leo18}
for CDW's, and find that the decrease follows a power-law form
as predicted \cite{privateLeo26}.
The transverse depinning occurs by the formation of 
longitudinal kinks. We do not observe a transverse depinning 
threshold for the perpendicular driven string.  Possible
experimental systems in which to investigate the predications in this work
include noise measurements and domain wall imaging for magnetic domains 
driven at increasing magnetic field sweep rates, 
as well as varied moving contact line speeds 
over a rough surface for increased drive. 
Our prediction for the reordering and 
the transverse barrier for the parallel driven string  
may be observable in polymers aligned with the
drive in disordered media or a single stream of liquid flowing over a rough 
surface at various longitudinal and transverse tilt angles. 

We thank M. Chertkov, T. Germann, L. Radzihovsky, and G. Zim{\' a}nyi
for useful discussions.  This work was supported by the US Department of
Energy under contract W-7405-ENG-36.

\end{document}